\documentclass[conference]{IEEEtran}

\usepackage[numbers,sort&compress]{natbib}
\usepackage{amsmath}
\usepackage{stmaryrd}
\usepackage{amssymb}
\usepackage{graphicx}
\usepackage{epstopdf}
\usepackage{stfloats}
\usepackage{dsfont}

\graphicspath{{fig/}}
\DeclareGraphicsExtensions{.eps}
\usepackage{epstopdf}
\usepackage{algorithm}
\usepackage{algorithmic}

\begin{document}
%
\title{Polar Coded HARQ Scheme with Chase Combining}
%
%
%
\author{\IEEEauthorblockN{Kai Chen, Kai Niu, Zhiqiang He and Jiaru Lin}
\IEEEauthorblockA{Key Laboratory of Universal Wireless Communications, Ministry of Education\\
Beijing University of Posts and Telecommunications, Beijing, China 100876\\
Email: \{kaichen, niukai, hezq, jrlin\}@bupt.edu.cn}}

%

\maketitle

\newtheorem{theorem}{Theorem}
\newtheorem{example}{Example}

\begin{abstract}

A hybrid automatic repeat request scheme with Chase combing (HARQ-CC) of polar codes is proposed.
The existing analysis tools of the underlying rate-compatible punctured polar (RCPP) codes for additive white Gaussian noise (AWGN) channels are extended to Rayleigh fading channels.
Then, an approximation bound of the throughput efficiency for the polar coded HARQ-CC scheme is derived.
Utilizing this bound, the parameter configurations of the proposed scheme can be optimized.
Simulation results show that, the proposed HARQ-CC scheme under a low-complexity SC decoding is only about $1.0$dB away from the existing schemes with incremental redundancy (\mbox{HARQ-IR}).
Compared with the polar coded \mbox{HARQ-IR} scheme, the proposed HARQ-CC scheme requires less retransmissions and has the advantage of good compatibility to other communication techniques.

\end{abstract}


\section{Introduction}
\IEEEPARstart{P}{olar} codes are the first structured codes that provably achieve the symmetric capacity of binary-input memoryless channels (BMCs) \cite{Arikan_first}.
Given a BMC $W$, after performing the channel transform, i.e., the channel combining and channel splitting operations, over a set of independent copies of $W$, a second set of synthesized channels is obtained.
As the transformation size goes to infinity, some of the resulting channels tend to be completely noised, and the others tend to be noise-free, where the fraction of the \mbox{noise-free} channels approaches the symmetric capacity of $W$.
By transmitting free bits over the noiseless channels and sending fixed bits over the others, polar coding with a very large code length $N$ can achieve the symmetric capacity under a successive cancellation (SC) decoder with both encoding and decoding complexity $O\left( N\log N \right)$.

In delay insensitive communications, hybrid automatic repeat request (HARQ) transmission scheme is widely used to obtain a capacity-approaching throughput efficiency \cite{Hagenauer_RCPC}, \cite{Rowitch_RCPT}, \cite{Yue_RCIRA}, \cite{rcpt_fading}.
There are mainly two types of HARQ schemes that are widely considered in practical systems.
One is Chase combining (HARQ-CC), where each retransmission block is identical to the original code block; and the other is incremental redundancy (HARQ-IR), where each retransmission consists of new redundancy bits from the channel encoder.
In \cite{Chen_harq}, an HARQ-IR scheme based on polar codes is proposed. The throughput performance is claimed to be as good as those based on LDPC and turbo codes with much lower decoding complexity.
Obviously, HARQ-IR has the potential of achieving better throughput compared to that with \mbox{HARQ-CC}.
However, HARQ-CC will have lower complexity than that with HARQ-IR.
That is because the use of IR requires some additional signaling (e.g., the retransmission numbers needs to be communicated to the receiver) and a much larger buffer is needed for IR.
Furthermore, since each retransmission is identical, it is much easier for HARQ-CC scheme to combine with other techniques, like coded modulation and space-time coding.

Therefore, this paper focuses on providing a polar coded HARQ-CC scheme.
As far as we know, this is the first \mbox{HARQ-CC} scheme based on polar codes.
The proposed scheme is applied to both additive white Gaussian noise (AWGN) channel and uncorrelated Rayleigh fast fading channel.
Given an information block with $K$ bits, the key problem of designing an HARQ-CC transmission scheme is to construct a rate-compatible punctured polar (RCPP) code with proper code length $N$, or equivalently, the code rate $R=\frac{K}{N}$.
The RCPP codes over AWGN channel are well studied in \cite{Niu_icc}.
Given an AWGN channel and a specific RCPP code, the block error rate (BLER) can be accurately predicted under the framework of channel polarization over parallel channels \cite{Chen_iet}.
The code construction and performance evaluation methods of RCPP codes are extended to the Rayleigh fading channels.
Utilizing these techniques, the code length $N$ can be optimized to maximize the throughput of the proposed HARQ scheme.

\begin{figure}[!t]
  \centering
  \includegraphics[width=0.95\columnwidth]{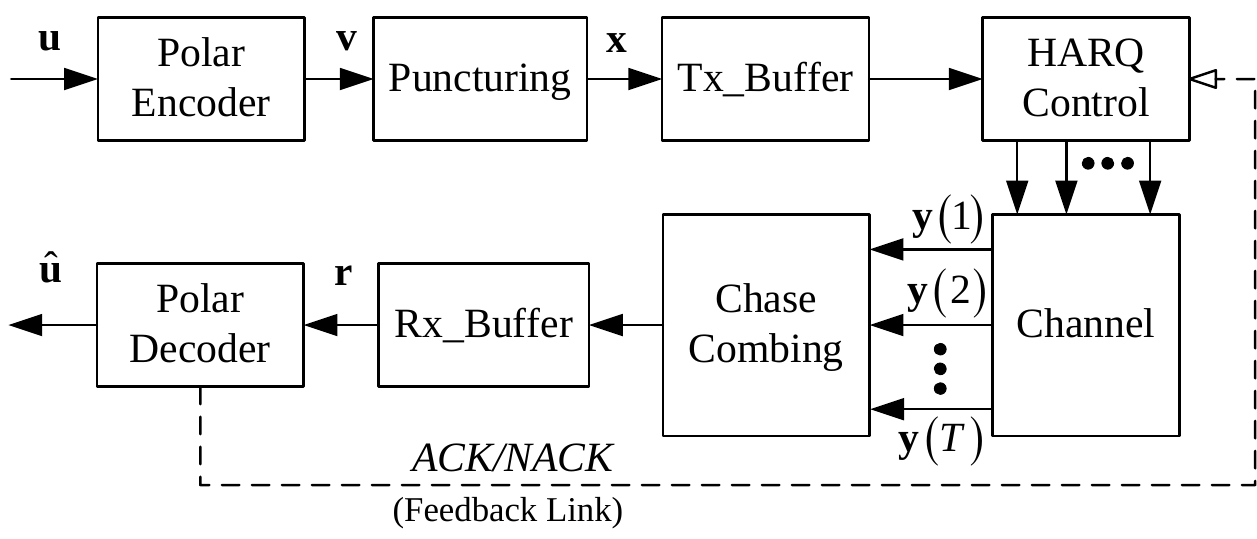}
  \caption{Block diagram of the proposed HARQ scheme.}
  \label{fig_diagram}
\end{figure}

The remaining sections of the paper are organized as follows.
\mbox{Section \ref{section_scheme}} gives a general description of the proposed scheme.
Section \ref{section_rcpp} reviews the underlying RCPP codes and extends the results to fading channels.
Section \ref{section_simulation} provides the simulation results of the proposed \mbox{HARQ-CC} scheme, and compares it with the existing HARQ-IR scheme based on polar codes, turbo codes and LDPC codes.
Finally, \mbox{Section \ref{section_conclusions}} concludes the paper.

\section{The Proposed Scheme}
\label{section_scheme}

This section gives an overall description of the proposed polar coded HARQ-CC transmission scheme.

\subsection{Notations}
We use calligraphic characters, such as $\mathcal{X}$, to denote sets.
Let $|\mathcal{X}|$ denote the number of elements in $\mathcal{X}$.
We write lowercase letters (e.g., $x$) to denote scalars, bold-face lowercase letters (e.g., $\textbf{x}$) to denote vectors, and $x_i$ to denote the $i$-th element of $\textbf{x}$.
For any $i \le j$, $\textbf{x}_{i:j}$ denotes a subvector of $\textbf{x}$, i.e., $\textbf{x}_{i:j}=({x}_{i}, {x}_{i+1}, \cdots, {x}_{j})$.
Throughout this paper, the base of the logarithm is $2$.

\subsection{Polar Coded HARQ-CC Transmission}
A source block $\textbf{u}$ which consists of $K$ information bits and $M-K$ frozen bits (usually are set to all-zero bits) is fed into a polar encoder, where $M$ is the code length of the base code and its value is restricted to some power of $2$.
The encoded sequence $\textbf{v}$ of $M$ bits is punctured into a punctured codeword $\textbf{x}$ of $N$ bits, $N\leq M$.
The mapping from the $K$-length information block to the $N$-length codeword is in fact an encoding procedure of a RCPP code \cite{Niu_icc}.
The block $\textbf{x}$ is buffered and sent over the channel.

At the receiver, the received signals of the $t$-th transmission and the corresponding \mbox{log-likelihood} ratios (LLR) are respectively written as $\textbf{y}(t)$ and $\textbf{r}(t)$, where $t=1,2,\cdots$ denotes the number of transmission trials.
The content of the buffer at the receiver is the combined LLRs $\textbf{r}$ of the received code bits.
After the first transmission, the content of the LLR buffer is initialized as $\textbf{r} \gets \textbf{r}(1)$.
The polar decoder tries to perform the decoding process based on $\textbf{r}$.
If the receiver fails to decode the codeword, i.e., the estimated source block $\hat{\textbf{u}}$ is not equal to $\textbf{u}$ which can be usually detected by a cyclic redundancy check (CRC) failure, an NACK (negative acknowledgement) is sent to the transmitter through the feedback channel.
And then, the $N$ encoded bits $\textbf{x}$ are retransmitted.
The new received signals $\textbf{y}(2)$ are translated to LLRs $\textbf{r}(2)$ and the LLRs of the first two transmission are combined, i.e., the content of the received LLR buffer is updated as $\textbf{r} \leftarrow \textbf{r}+\textbf{r}(2)$, where the operation ${+}$ of two vectors denotes the termwise addition.
The polar decoder tries to decode $\hat{\textbf{u}}$ according to the updated $\textbf{r}$.
This process continues until the transmitter receives an ACK (acknowledgement), or a maximum number of permitted transmissions $T$ is achieved.

A block diagram of the proposed HARQ-CC scheme is shown in \mbox{Fig. \ref{fig_diagram}}.

\subsection{Channel Model}

Without loss of generality, only binary phase shift keying (BPSK) is considered in this paper.
When transmitting a RCPP codeword $\textbf{x}$ over the channel, the receiving signals at the $t$-th transmission $\textbf{y}(t)$ are as follows:
\begin{equation}
\label{equ_signal}
y_i(t) = a_i(t) \cdot s_i + z_i(t)
\end{equation}
where $t \in \{1,2,\cdots, T\}$, $i \in \{1, 2, \cdots, N\}$, $s_i=1-2 x_i$ is the signal after BPSK modulation, $z_i(t)$ is the Gaussian noise with zero mean and variance $\sigma^2$, i.e., $z_i \sim N(0, \sigma^2)$ and $a_i(t)$ is the fading factor with average power gain $\mathbb{E}[a^2_i(t)]=1$.

During the whole transmission procedure, the noise variance $\sigma^2$ of the Gaussian noise is supposed to be constant and is known to both the transmitter and receiver.
However, the instant values of $a_i$ are only available at the receiver, while the transmitter only has a prior knowledge of the probability distribution function (PDF) of the fading factor.

In this paper, the proposed scheme is applied to the transmissions over AWGN channel and uncorrelated Rayleigh fast fading channel.

\subsubsection{AWGN Channel}

When the fading factor \mbox{$a_i(t)=1$} for all $i\in\{1,2,\cdots, N\}$ and $t\in \{1,2,\cdots,T\}$, the signal model defined in equation (\ref{equ_signal}) degrades into a transmission over a binary-input AWGN channel.

The symmetric capacity of a binary-input AWGN channel $W$ with noise variance $\sigma^2$ is \cite{IT_book}
\begin{equation}
\label{equ_agwncap}
I_{G}(\sigma^2) = -\int_{-\infty}^{+\infty}{p(y) \cdot \log p(y)\text{d}y} - \frac{1}{2}\log{2\pi e \sigma^2 }
\end{equation}
where
\begin{equation}
p(y)=\frac{1}{2\sqrt{2\pi \sigma^2}}\left(e^{\frac{-(y-1)^2}{2\sigma^2}}+e^{\frac{-(y+1)^2}{2\sigma^2}}\right)
\end{equation}

\medskip

\subsubsection{Uncorrelated Rayleigh Fast Fading Channel}

In this scenario, for $i\in\{1,2,\cdots, N\}$ and $t\in \{1,2,\cdots,T\}$, all the $a_i(t)$ in (\ref{equ_signal}) are i.i.d. and are with PDF
\begin{equation}
\label{equ_rpdf}
p(a) = 2a \exp\left({-a^2}\right)
\end{equation}

Given a Rayleigh fast fading channel $W$ with noise variance $\sigma^2$, the ergodic capacity of $W$ can be calculated as
\begin{equation}
\label{equ_fadingcap}
I_{R}(\sigma^2) = \int _{0}^{+\infty}{I_{G}\left({\sigma}^2/{a^2}\right)p(a)\text{d}a}
\end{equation}

\medskip

\section{Rate-Compatible Punctured Polar Codes}
\label{section_rcpp}

The proposed HARQ transmission scheme is based on the RCPP codes introduced in \cite{Niu_icc}.

Similar to constructing a conventional polar code, after performing the channel transform over $M=2^{\lceil \log N \rceil}$ independent uses of the original channel $W$, where $\lceil \cdot \rceil$ is the ceiling function, we get ${M}$ successive uses of synthesized binary-input channels $W_{M}^{(i)}$, $i=1,2,\cdots,M$.
Given a symmetric BMC $W$, let $\textsf{a}$ denote the probability density function (PDF) of the log-likelihood ratio (LLR) of the received bit when a bit zero is transmitted.
The reliability of $W$ can be measured as the error probability
\begin{equation}
\label{equ_bitpe}
P_e(W) = \int \nolimits_{-\infty}^{0} \textsf{a}(z) \text{d}z
\end{equation}
Let $\textsf{a}_{M}^{(i)}$, $i=1,2,\cdots,M$ denote the LLR PDFs of the received bit from $W_{M}^{(i)}$ when all-zero information bits are transmitted.
After calculating $\textsf{a}_{M}^{(i)}$ by density evolution (DE) \cite{Mori_DE}, the reliabilities of $W_{M}^{(i)}$ are determined by (\ref{equ_bitpe}).
In transmitting a binary information block of $K$ bits, the $K$ most reliable polarized channels $W_{M}^{(i)}$ with indices $i \in \mathcal{A}$ are selected to carry the $K$ information bits, where $\mathcal{A} \subset \{1, 2, \cdots, M\}$ and $|\mathcal{A}|=K$, and these channels are called \emph{information channels}; and the others are called \emph{frozen channels} and are used to transmit a fixed sequence.

Different from the conventional polar codes, $M-N$ output bits of the polar encoder should be punctured when dealing with a RCPP code.
Therefore, before performing the channel transform, the underlying channel uses corresponding to these punctured bits should be replaced by virtual channels \cite{Chen_iet}, which have the same input and output alphabets as $W$ but with zero capacities.
As for determining the positions of the punctured bits, without loss of generality, the quasi-uniform puncturing scheme in \cite{Niu_icc} is adopted in this paper which is claimed to be an efficient and empirical good solution.
The punctured positions are represented by an $M$-dimensional binary vector $\textbf{p}$ (which is called puncturing pattern), where the $0$s indicate the positions of the punctured bits and $1$s indicate the positions of the reserved bits.
Given the length of base code $M$ and the length of punctured code $N$, the puncturing pattern $\textbf{p}$ can be determined as follows:
\begin{algorithm}[!ht]
\caption{Determine the Puncturing Pattern \cite{Niu_icc} }
\renewcommand{\algorithmicrequire}{\textbf{Input:}}
\renewcommand{\algorithmicensure}{\textbf{Output:}}
\label{alg_punc_pattern}
\begin{algorithmic}[1]
\REQUIRE Code length of the base code $M$;\\
\quad \:\:Code length of the punctured code $N$;
\ENSURE Puncture pattern \textbf{p};\\
\STATE Initialize $\textbf{p}$ as a $M$-length all one vector, i.e., for all \mbox{$i \in \{1, 2, \cdots, M\}$, set $p_i \leftarrow 1$};
\STATE Set the first $M-N$ elements of $\textbf{p}$ as zeros, i.e., for \mbox{$i \in \{1, 2, \cdots, M-N\}$, set $p_i \leftarrow 0$};
\STATE Perform the bit-reversal permutation on $\textbf{p}$.
\RETURN $\textbf{p}$;
\end{algorithmic}
\end{algorithm}

\noindent More details of RCPP codes can be found in \cite{Niu_icc}.

Similar to the conventional polar codes \cite{Arikan_first}, RCPP codes can also be decoded using SC decoding algorithm.
The BLER of a RCPP code under SC decoding can be evaluated as
\begin{equation}
\label{equ_bler}
P_{B}(N, K, M, \mathcal{A}) = \sum \nolimits _{i\in\mathcal{A}}{P_e\left({W}_{M}^{(i)}\right)}
\end{equation}
Note that, the performance of the RCPP codes relies heavily on the puncturing patterns.
However, only the puncturing scheme in Algorithm \ref{alg_punc_pattern} is considered in this paper, and the puncturing pattern $\textbf{p}$ can be uniquely determined by $M$ and $N$.
Therefore, $P_{B}$ in (\ref{equ_bler}) does not involve a $\textbf{p}$ in the parameter list.

In the case of AWGN channel, the BLER performance under SC decoding (\ref{equ_bler}) can be evaluated efficiently by Gaussian approximation (GA) of DE \cite{Trifonov_GA}.
As shown by the results in \cite{Trifonov_GA}, the estimated BLER in (\ref{equ_bler}) obtained by GA is very accurate in practical signal-to-noise ratio (SNR) regimes.

\begin{figure}[!t]
  \centering
  \includegraphics[width=0.8\columnwidth]{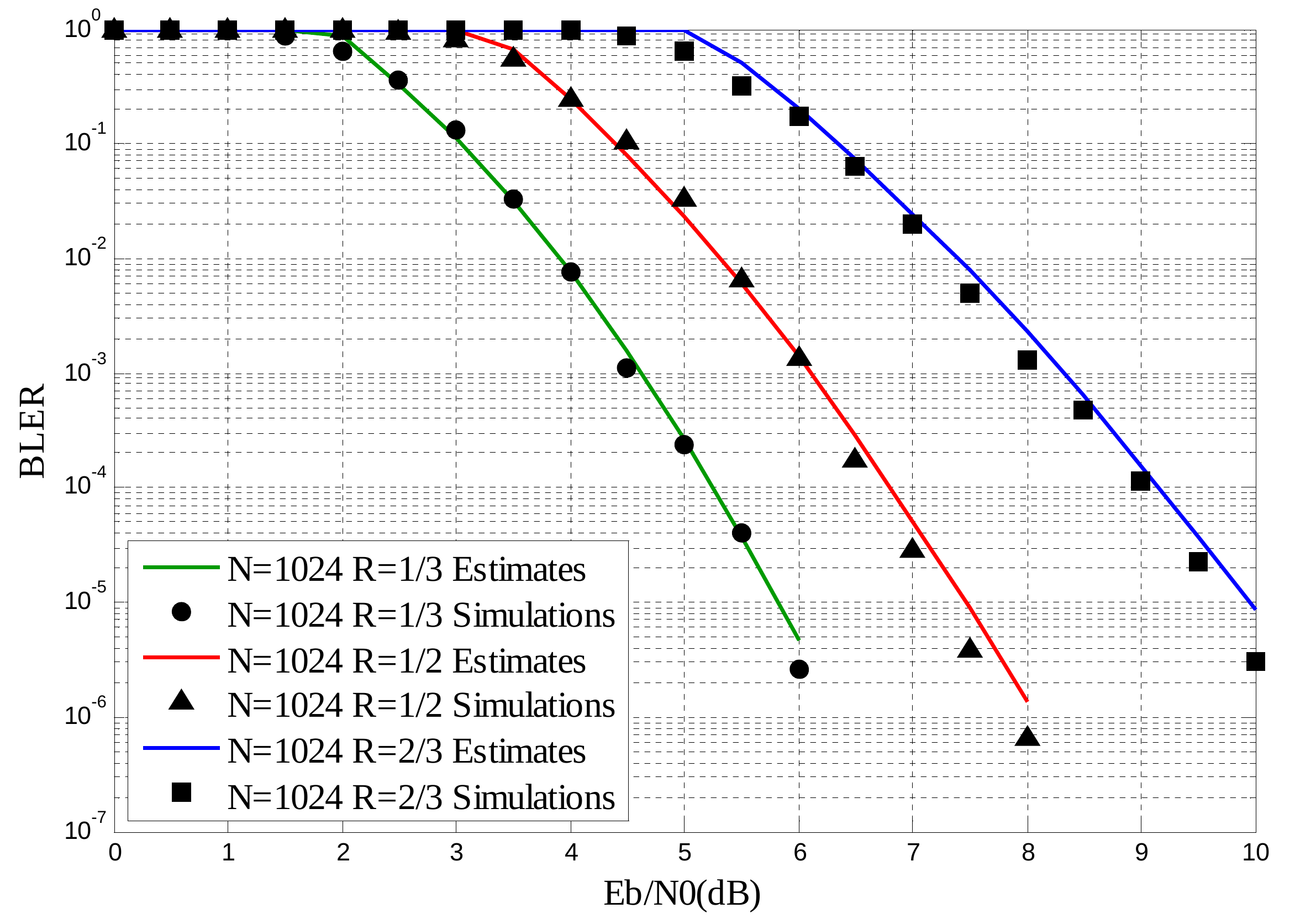}
  \caption{Performance of RCPP codes under SC decoding over binary-input uncorrelated Rayleigh fast fading channels and the corresponding BLER bound obtained by GA via equivalent binary-input AWGN channels.}
  \label{fig_gafading}
\end{figure}

In the case of uncorrelated Rayleigh fast fading channel, since the instant fading factors are not available at the receiver, the existing construction method of RCPP codes cannot be employed directly.
In this paper, we propose to construct RCPP codes by approximating the fading channel $W$ with $\sigma^2$ using an AWGN channel $W_{eq}$ with $\sigma^2_{eq}$,  where the capacity of $W_{eq}$ equals to the ergodic capacity of $W$, i.e.,
\begin{equation}
I_G(\sigma^2_{eq}) = I_{R}(\sigma^2)
\end{equation}
where $I_G$ and $I_R$ are calculated as (\ref{equ_agwncap}) and (\ref{equ_fadingcap}), respectively.
The code construction and performance evaluation is then performed over the equivalent AWGN channel $W_{eq}$ in the same way as that of the AWGN case.
The BLER performances over Rayleigh fading channels of a set of RCPP codes with $N=1024$ and the corresponding bounds (\ref{equ_bler}) obtained by the equivalent AWGN channels are shown in Fig.\ref{fig_gafading}.
The bounds and the simulation curves are matched quite well.

\section{Design an HARQ-CC Scheme of Polar Codes}

Designing an optimal HARQ-CC scheme is equivalent to constructing a RCPP code that can maximize the throughput efficiency.
This section first gives an approximation bound of the throughput efficiency under a specific configuration of the underlying RCPP code, then the construction algorithm of the proposed HARQ-CC scheme is described in detail.

\subsection{An Approximation Bound of Throughput Efficiency}

Similar to the HARQ-IR scheme in \cite{Chen_harq}, the throughput efficiency of a specific HARQ-CC scheme can also be estimated by an approximation bound.

To transmit an information block of $K$ bits with the proposed HARQ-CC scheme which allows at most $T$ transmissions, we need to search for the optimal code length of the underlying RCPP code.
After $t$ transmissions, a total of $t$ (noised) copies of the codeword $\textbf{x}$ are received from the channel.
Let $E_t$ with $t=1,2,\cdots, T$ denote the event that the information block \emph{cannot} be correctly decoded after the first $t$ transmissions, and $\overline{E_t}$ denote the complementary event of $E_t$.
We write $\Pr({E_t})$ to denote the probability of event $E_t$.
Particularly, we write $E_0$ to denote the event that the information block cannot be decoded by the receiver before transmitting any bits. Obviously, $\Pr(E_0)=1$.

When transmitting information blocks of $K$ bits, the average numbers of the successfully received information bits $\mathbb{E}[K]$ is
\begin{eqnarray}
\mathbb{E}\left[ K \right]
\label{equ_EK}
&=&K\cdot \left( 1-\Pr\left( {E_T} \cap {E_{T-1}} \cdots \cap {E_{0}} \right) \right)
\end{eqnarray}
and the total transmitted bits $\mathbb{E}[N]$ is
\begin{eqnarray}
\mathbb{E}\left[N\right] &=& \sum \limits _{t=1}^{T}{N \cdot \Pr( \overline{E_t} \cap {E_{t-1}} \cap {E_{t-2}} \cdots \cap {E_{0}})}\nonumber \\
\label{equ_EN}
                &+& N \cdot \Pr( {E_T} \cap {E_{T-1}} \cdots \cap {E_{0}})
\end{eqnarray}

Then, the throughput efficiency can be written as
\begin{equation}
\label{equ_throughput}
\eta = \frac{\mathbb{E}[K]}{\mathbb{E}[N]}
\end{equation}

Obviously, we have
\begin{eqnarray}
\label{equ_a}
\Pr( {E_t} \cap  \cdots \cap {E_{0}}) \le \Pr(E_t)
\end{eqnarray}

Similar to that in \cite{Chen_harq}, we would like to use the following \mbox{approximation}
\begin{eqnarray}
\label{equ_hypothesis}
&&\Pr( \overline{E_t} \cap \cdots \cap {E_{0}}) \approx \Pr({{E}_{t-1}})-\Pr({{E}_{t}})
\end{eqnarray}

After the $t$-th transmission over AWGN channel with $\sigma^2$, the received LLR vector $\textbf{r}$ after Chase combing is equivalent to that received after one transmission over an AWGN channel with Gaussian noise variance $\sigma^2 / t$.
When transmitting over the Rayleigh fading channels, the problem will be much more complex because the equivalent fading factor after Chase combining is no longer Rayleigh distributed, and the PDF of the equivalent fading factor is in the form of $t$ self-convolutions of (\ref{equ_rpdf}).
For the ease of performance evaluation, we always use equivalent AWGN channels to approximate the fading channels when constructing the RCPP codes.
After $t$ transmissions, the decoding is performed based on the combined LLRs of $t$ (noised) copies of the identical codeword $\textbf{x}$ received from the channel.
Thus, when the channel is with Gaussian noise variance $\sigma^2$, $\Pr({E_t})$ in (\ref{equ_a}) and (\ref{equ_hypothesis}) is in fact the BLER of the RRCP code when transmitting over an equivalent AWGN channel with an noise variance $\sigma^2 /t$ (or $\sigma^2_{eq}/t$ for fading channel).
So, for both scenarios of AWGN and Rayleigh fading channels, the values of $\Pr({E_t})$ can be efficiently evaluated by (\ref{equ_bler}) using GA as introduced in \mbox{section \ref{section_rcpp}}.

Therefore, the throughput efficiency in (\ref{equ_throughput}) can be approximately calculated as
\begin{eqnarray}
\label{equ_eta}
\eta &\approx& \frac{K \cdot (1-\Pr(E_T))}{ \sum \nolimits_{t=1}^{T}{ N \cdot (\Pr(E_{t-1})-\Pr(E_t)) } + N \cdot \Pr(E_T) } \nonumber \\
&=&\frac{K \cdot (1-\Pr(E_T))}{ N \cdot \left(1+\sum \nolimits_{t=1}^{T-1}{ \Pr(E_t)} \right)}
\end{eqnarray}

Since the substitution of $\Pr(E_t)$ for $\Pr( {E_t} \cap  \cdots \cap {E_{0}})$ in (\ref{equ_a}) is an upper bound, and the approximation $\Pr(E_{t-1})-\Pr(E_t)$ for $\Pr( \overline{E_t} \cap {E_{t-1}} \cap \cdots \cap {E_{0}})$ in (\ref{equ_hypothesis}) is usually also an upper bound, the approximation of the throughput efficiency in (\ref{equ_eta}) tends to be a lower bound of $\eta$.

\subsection{Searching for the Optimal HARQ-CC Scheme}
\label{section_harq:design}
Utilizing (\ref{equ_eta}), an HARQ-CC scheme with information block size $K$ can be constructed via a greedy search.
Given the length of the (punctured) codeword $N$, the code length of the base code $M$ is restricted to the least available value that is larger than $N$, i.e, $M=2^{\log \lceil N \rceil}$, and the puncturing pattern is determined by Algorithm \ref{alg_punc_pattern}.
The information channel indices $\mathcal{A}$ of the RCPP code are selected to minimize the BLER of the first transmission attempt.
The BLERs after $t$ transmissions with $t=1,2,\cdots, T$ can be evaluated by (\ref{equ_bler}) using GA.
Then, the throughput efficiency can be estimated by (\ref{equ_eta}).
All the potential configurations of the code length $N$ taking values from $K$ to $\lfloor Q/T \rfloor$ are checked, where $Q$ is the number of permitted transmitted bits during the entire transmission procedure and $\lfloor \cdot \rfloor$ is the floor function.
Finally, the optimal configuration of the code length $N$ with the highest throughput efficiency is recorded.

The search algorithm is summarized in \mbox{Algorithm \ref{alg_harq}}.
The inputs include the information block length $K$, the number of the permitted transmitted bits $Q$, the maximum number of transmission trials $T$ and the variance of Gaussian noise $\sigma^2$ ( $\sigma^2_{eq}$ for the case of Rayleigh fading channel).
The algorithm outputs the optimal (punctured) code length $N$.

\begin{algorithm}[!ht]
\caption{Design a Polar Coded HARQ-CC Scheme}
\renewcommand{\algorithmicrequire}{\textbf{Input:}}
\renewcommand{\algorithmicensure}{\textbf{Output:}}
\label{alg_harq}
\begin{algorithmic}[1]
\REQUIRE Information block length $K$;\\
\quad \: Maximum number of transmission trials $T$; \\
\quad \: Number of permitted transmitted bits $Q$; \\
\quad \: Variance of Gaussian noise $\sigma^2$ ($\sigma^2_{eq}$ for Rayleigh fading case);
\ENSURE  Length of the punctured codeword $N$;
\STATE Initialize the $N \gets 0$, $M \gets 0$, and the optimal throughput efficiency $\eta_{opt} \gets 0$;
\FOR{$n \gets K:\lfloor Q/T \rfloor$}
    \IF{$\frac{K}{n}<\eta_{opt}$}
    \STATE Terminate the searching loop;
    \ENDIF
    \STATE The length of the base code is set as $m \gets 2^{\lceil \log\left(n\right) \rceil}$;
    \STATE Construct a set of information channel indices $\mathcal{A}$ under the channels with parameter $\sigma^2$.
    \STATE Allocate a temporary $T$-dimensional vector $\textbf{q}$;
    \FOR{$t \gets 1:T$}
        \STATE Estimate the error probability after $t$ transmissions, i.e., $q_t \gets P_B(n, K, m, \mathcal{A})$, where the underlying channel is with parameter $\sigma^2 /t$;
    \ENDFOR
    \STATE Calculate the throughput of the temporary scheme:
        \begin{eqnarray}
        \eta=\frac{K \cdot (1-q_T)}{ n \cdot \left(1+\sum \nolimits_{t=1}^{T-1}{ q_t} \right)}
        \end{eqnarray}
    \IF{$\eta > \eta_{opt}$}
        \STATE Record the optimal code length $N\gets n$;
        \STATE Update the optimal throughput efficiency $\eta_{opt} \gets \eta$;
    \ENDIF
\ENDFOR
\RETURN $N$;
\end{algorithmic}
\end{algorithm}

In \mbox{Algorithm \ref{alg_harq}}, the outer loop at line 2 is executed $\lfloor Q/T \rfloor-K+1$ times.
An early termination is given in line 5 to \mbox{line 7}: if the potential code length $n$ is too large to get the higher throughput than the already obtained optimal configuration, the search procedure is then terminated.
The most expensive operations are the BLER estimations in line 12 to line 14.
According to \cite{Trifonov_GA}, the complexity of evaluating (\ref{equ_bler}) using GA is $O\left(N\log N\right)$.
Taking the outer loop of $n$ into account, the overall complexity of \mbox{Algorithm \ref{alg_harq}} is upper bounded by $O\left( \left( {{Q}^{2}}/T-Q K \right) \log \left( Q/T \right) \right)$.

In comparison, the construction complexity of HARQ-IR of polar codes is claimed to be $O\left( Q^2\log Q\right)$ \cite{Chen_harq}.
Note that, under the context of HARQ-CC, $Q$ is at least $T$ times larger than $N$.
Therefore, the construction complexity of the proposed polar coded HARQ-CC scheme is much lower than that of the HARQ-IR in \cite{Chen_harq}.

\section{Simulation Results}
\label{section_simulation}
In this section, the performance of the proposed polar coded \mbox{HARQ-CC} scheme is evaluated via simulations over AWGN and Rayleigh fading channels.
All the RCPP codes are with information block lengths $K=1024$ and decoded by SC algorithm.
The proposed \mbox{HARQ-CC} schemes are constructed with the number of permitted transmitted bits $Q=16384$ and maximum number of transmission trials $T=6$.

\begin{figure}[!t]
  \centering
  \includegraphics[width=0.8\columnwidth]{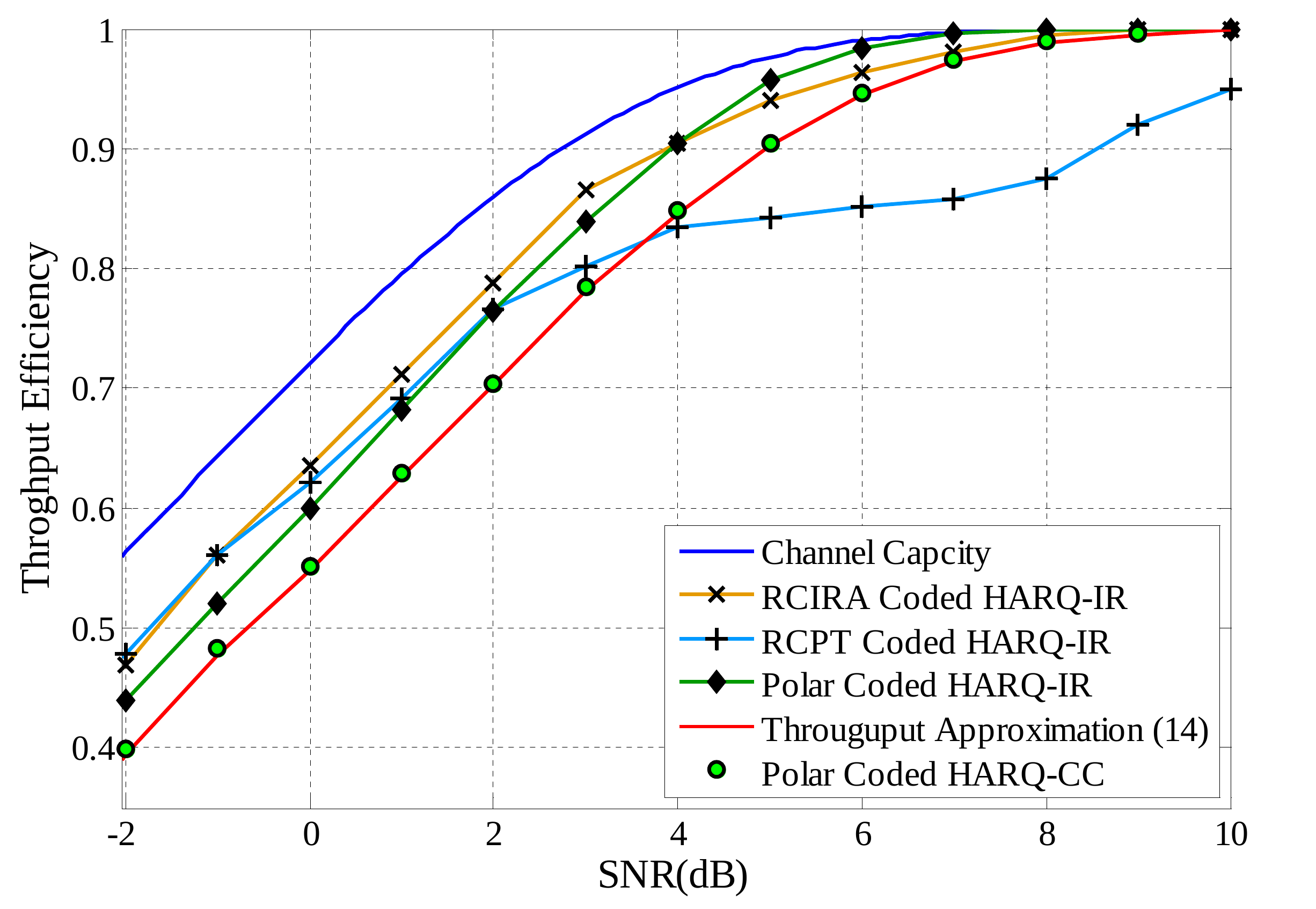}
  \caption{Throughput efficiency of HARQ transmission schemes in AWGN channels, where the polar codes are with $K=1024$, and results for RCIRA codes are from \cite{Yue_RCIRA} ($K=512$) and RCPT codes are from \cite{Rowitch_RCPT} ($K=1024$).}
  \label{fig_sim_awgn}
\end{figure}

Fig. \ref{fig_sim_awgn} shows the throughput efficiency of the proposed HARQ-CC scheme over \mbox{BI-AWGNCs}, the configurations are shown in Table \ref{tab_config_awgn}.
The simulation results of the proposed HARQ-CC scheme and the approximation (\ref{equ_eta}) are well matched.
In comparison, the throughput efficiency curves of the \mbox{HARQ} schemes based on the rate-compatible punctured turbo codes (\mbox{RCPT}) \cite{Rowitch_RCPT} and the rate-compatible irregular repeat-accumulate (\mbox{RCIRA}) codes \cite{Yue_RCIRA} (as a representative class of LDPC codes) are provided.
Moreover, the performance of HARQ-IR of polar codes in \cite{Chen_harq} is also provided.

\begin{table}[!t]
\centering
\caption{Configurations of HARQ-CC over AWGN channels with $K=1024$, $Q=16384$ and $T=6$}
\label{tab_config_awgn}
\begin{tabular}{c|c||c|c||c|c}
\hline
\hline
SNR(dB)   &   $N$   &   SNR(dB)   &   $N$   &   SNR(dB)   &   $N$    \\
\hline
$-2.00$   &   $2489$   &   $3.00$   &   $1280$   &   $8.00$   &   $1032$    \\
$-1.00$   &   $2048$   &   $4.00$   &   $1184$   &   $9.00$   &   $1027$    \\
$0.00$   &   $1808$   &   $5.00$   &   $1120$   &   $10.00$   &   $1025$    \\
$1.00$   &   $1583$   &   $6.00$   &   $1072$   &      &       \\
$2.00$   &   $1420$   &   $7.00$   &   $1046$   &      &       \\
\hline
\hline
\end{tabular}
\end{table}

As the figure shows, the proposed scheme works a little worse than the HARQ-IR schemes of turbo codes and LDPC codes.
In the low SNR regime, the performance loss of the polar coded HARQ-CC against turbo/LDPC coded HARQ-IR is about  $1.0$dB.
As the SNR goes higher, the performance gap between the proposed scheme and turbo/LDPC coded schemes becomes smaller.
When the SNR is above $4.0$dB, the proposed scheme achieves better throughput efficiency than that of turbo coded scheme.
One reason for the performance advantage in high SNR regimes is that the code length of the RCPP codes can be adjusted precisely in a step of only $1$ bit.  Utilizing the accurate BLER and throughput bounds, the polar coded schemes can be well optimized.
While the performance of the turbo code is difficult to evaluate, the choices for candidate code rates of turbo codes are usually restricted to a size-limited set.
It is hard to optimize the turbo coded scheme in both low and high SNR regimes.

Moreover, the performances of the polar coded HARQ schemes provided in this paper are obtained under the SC decoding. It is shown in \cite{Niu_CADec} that the performance loss of polar codes under SC is greater than $1.0$dB compared with the turbo codes and LDPC codes, but the complexity of the SC decoding is much lower than that of the \mbox{log-MAP} decoding for turbo codes or the belief propagation decoding for LDPC codes.
As shown in \cite{Tal_List} \cite{Niu_CADec}, the inherently embedded CRC bits in HARQ schemes can be utilized to greatly improve the BLER performance of the polar codes.
By properly configuring the decoder, the enhanced polar decoding scheme will be no more complex than those of turbo/LDPC codes, while the polar codes can achieve a better BLER performance.
If these enhanced decoding schemes are applied, the throughput performance of the proposed polar coded HARQ-CC scheme can be further improved.
However, it lacks a proper bound for the performance of polar codes under CRC aided decoding schemes.
The probabilities $\Pr(E_t)$ in (\ref{equ_hypothesis}) are no longer easy to evaluate as that under SC decoding.
We would like to leave this problem for future research.

\begin{table}[!t]
\centering
\caption{BLER at the First Transmission and the Average Number of Transmissions of Polar Coded HARQ-CC and HARQ-IR over AWGN channels with $K=1024$, $Q=16384$ and $T=6$}
\label{tab_comparison}
\begin{tabular}{l||c|c||c|c}
\hline
\hline
&\multicolumn{2}{|c||}{HARQ-CC} & \multicolumn{2}{c}{HARQ-IR} \\
\hline
SNR(dB)	& $\Pr(E_1)$	& Avg. Trans.  & $\Pr(E_1)$	& Avg. Trans. \\
\hline
-2.00	    & 4.18e-002	& 1.042  & 3.38e-001	& 1.766\\
0.00	    & 3.23e-002	& 1.032  & 3.80e-001	& 1.831\\
2.00	    & 2.81e-002	& 1.028  & 4.61e-001	& 1.973\\
4.00	    & 2.24e-002	& 1.022  & 4.09e-001	& 1.791\\
6.00	    & 1.11e-002	& 1.011  & 6.21e-001	& 2.124\\
8.00	    & 4.71e-003	& 1.005  & 2.85e-001	& 1.431\\
\hline
\hline
\end{tabular}
\end{table}

Also shown in Fig. \ref{fig_sim_awgn}, not much surprisingly, the proposed polar coded HARQ-CC scheme suffers a slight performance deterioration compared with the polar coded \mbox{HARQ-IR} scheme, where the latter benefits from additional coding gains.
A more detailed comparison between these two schemes on the BLER at the first transmission (i.e., $\Pr(E_1)$) and the average required transmission numbers are given in \mbox{Table \ref{tab_comparison}}.
The HARQ-CC scheme prefers to transmit the codeword more reliably at the first transmission, thus it requires less transmissions; while the length of the retransmission blocks of the HARQ-IR scheme is variable-sized, one or two times of retransmissions may not make significant deterioration to the overall throughput, so the value of $\Pr(E_1)$ tends to be relatively larger and it requires more transmissions in average.
The requirement for more transmissions is equivalent to a higher overhead on the feedback link and a longer temporal delay to successively transmitting a information block.
This is one of the advantages of the proposed HARQ-CC against the existing polar coded HARQ-IR scheme.

\begin{figure}[!t]
  \centering
  \includegraphics[width=0.8\columnwidth]{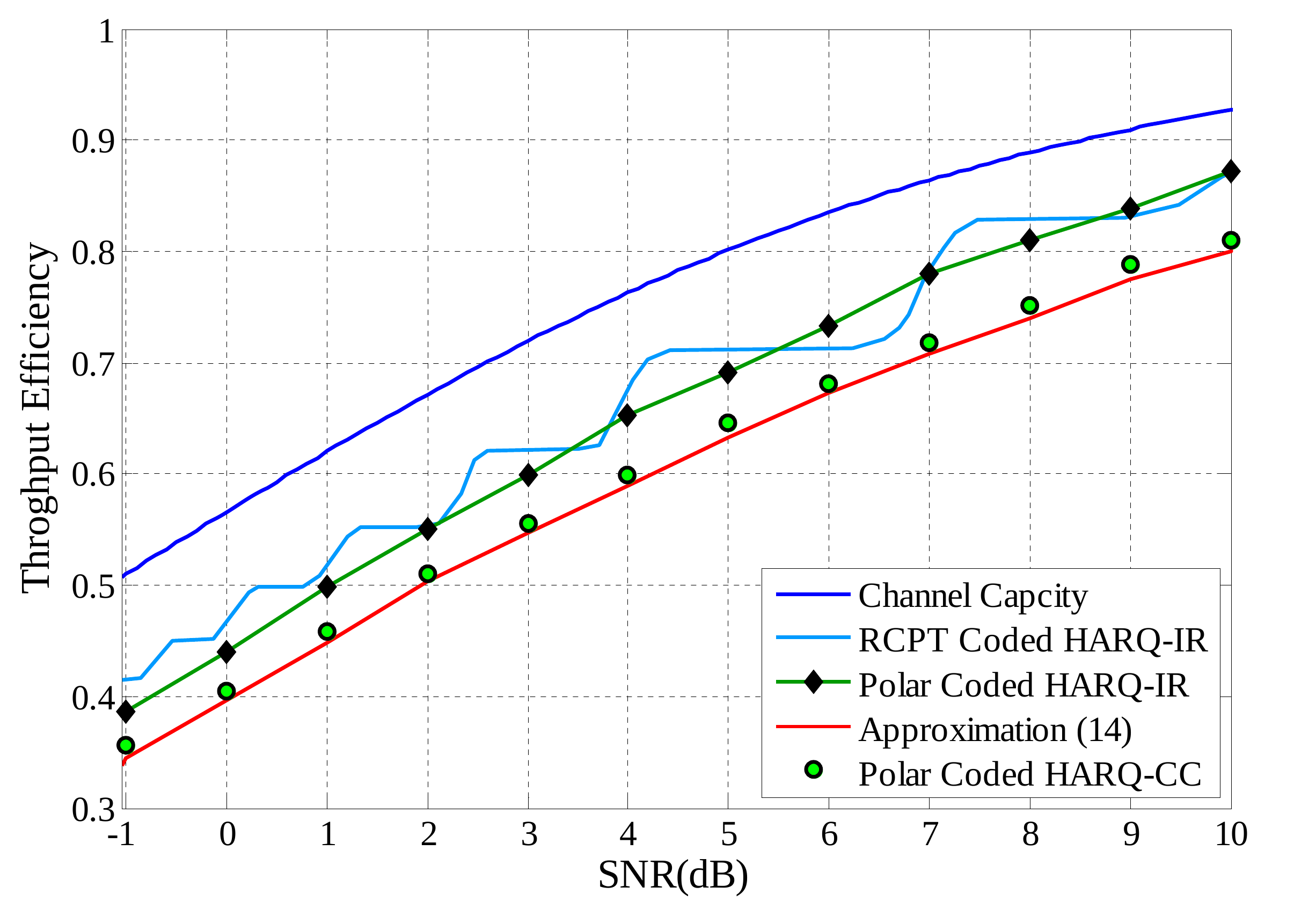}
  \caption{Throughput efficiency of HARQ transmission schemes in Rayleigh fading channels, where the polar codes are with $K=1024$, the results for RCPT codes are from \cite{rcpt_fading} ($K=5000$).}
  \label{fig_sim_fading}
\end{figure}

Similar simulations are also performed over the uncorrelated Rayleigh fast fading channels.
The throughput performance and configurations of the proposed HARQ schemes are given in Fig. \ref{fig_sim_fading} and Table \ref{tab_config_fading}, respectively.
The simulation results show that the approximation bound (\ref{equ_eta}) is also quite tight when transmitting over the Rayleigh fading channel.
Like the AWGN case, similar conclusions can be drawn.


Compared with the existing polar coded HARQ-IR scheme, HARQ-CC suffers from a slight performance deterioration.
However, since the retransmitted blocks under HARQ-CC scheme are identical to the initial one, the proposed scheme is more compatible to the new transmission techniques such as polar coded modulation \cite{Seidl_PCM}.

\begin{table}[!t]
\centering
\caption{Configurations of HARQ-CC over Uncorrelated Rayleigh fast fading channels with $K=1024$, $Q=16384$ and $T=6$}
\label{tab_config_fading}
\begin{tabular}{c|c||c|c||c|c}
\hline
\hline
SNR(dB)   &   $N$   &   SNR(dB)   &   $N$   &   SNR(dB)   &   $N$    \\
\hline
$-1.00$   &   $2730$   &   $3.00$   &   $1819$   &   $7.00$   &   $1408$    \\
$-0.00$   &   $2472$   &   $4.00$   &   $1682$   &   $8.00$   &   $1344$    \\
$1.00$   &   $2180$   &   $5.00$   &   $1552$   &   $9.00$   &   $1280$    \\
$2.00$   &   $1980$   &   $6.00$   &   $1488$   &   $10.00$   &   $1254$    \\
\hline
\hline
\end{tabular}
\end{table}

\section{Conclusions}
\label{section_conclusions}

An HARQ-CC scheme of polar codes is proposed.
As far as we know, this is the first \mbox{HARQ-CC} scheme based on polar codes.
Simulation results show that, the proposed scheme is only about $1.0$dB away from the existing polar coded HARQ-IR scheme.
But the new proposed polar coded HARQ-CC scheme requires less retransmissions and has the advantage of good compatibility to other transmission techniques.

\section*{Acknowledgment}
This work was supported by
the National Natural Science Foundation of China (No. 61171099),
the National Science and Technology Major Project of China (No. 2012ZX03003-007, No. 2012ZX03004-005-002, and No. 2013ZX03003012) and Qualcomm \mbox{Corporation}.

\ifCLASSOPTIONcaptionsoff
  \newpage
\fi

\end{document}